\documentclass[aps,prl,twocolumn,superscriptaddress]{revtex4}
\usepackage{graphicx}
\usepackage{amssymb}
\usepackage{pifont}
\begin{document}

%\preprint{}


\title{Strong Coupling between Antiferromagnetic and Superconducting Order Parameters in CeRhIn$_5$ Studied by In-NQR Spectroscopy}

\author{M.~Yashima}
\affiliation{Department of Materials Engineering Science, Osaka University, Osaka 560-8531, Japan}
\author{H.~Mukuda}
\affiliation{Department of Materials Engineering Science, Osaka University, Osaka 560-8531, Japan}
\author{Y.~Kitaoka}
\affiliation{Department of Materials Engineering Science, Osaka University, Osaka 560-8531, Japan}
\author{H.~Shishido}
\altaffiliation{Present Address: Department of Physics, Kyoto University, Kyoto 606-8502, Japan}
\affiliation{Department of Physics, Graduate School of Science, Osaka University, Osaka 560-0043, Japan}
\author{R.~Settai}
\affiliation{Department of Physics, Graduate School of Science, Osaka University, Osaka 560-0043, Japan}
\author{Y.~\=Onuki}
\affiliation{Department of Physics, Graduate School of Science, Osaka University, Osaka 560-0043, Japan}
\affiliation{Advanced Science Research Center, Japan Atomic Energy Research Institute, Tokai, Ibaraki 319-1195, Japan}

%\thanks{}
%\affiliation{}
\begin{abstract}
We report on a novel pressure ($P$)-induced  evolution of magnetism and superconductivity (SC) in a helical magnet CeRhIn$_5$ with an incommensurate wave vector $Q_i=(\frac{1}{2},\frac{1}{2},0.297)$ through the $^{115}$In nuclear quadrupole resonance (NQR) measurements under $P$. Systematic measurements of the $^{115}$In-NQR spectrum reveal that the commensurate antiferromagnetism (AFM) with $Q_c=(\frac{1}{2},\frac{1}{2},\frac{1}{2})$ is realized above $P_m \sim$ 1.7 GPa. An important finding is that the size of SC gap and $T_c$ increase as the magnitude of the AFM moment decreases in the $P$ region, where SC uniformly coexists with the commensurate AFM. This result provides evidence of strong coupling between the commensurate AFM order parameter (OP) and SC OP.
\end{abstract}
\vspace*{5mm}

\maketitle
\section{I. INTRODUCTION}

Various studies on the Ce115 family have revealed an intimate relationship between antiferromagnetism (AFM) and superconductivity (SC). CeRhIn$_5$ is a heavy-fermion (HF) helical magnet with $Q_i=(\frac{1}{2},\frac{1}{2},0.297)$ and the N\'{e}el temperature ($T_N$) = 3.8 K at the ambient pressure ($P$ = 0), as shown in Fig.1a \cite{Bao}, and it exhibits $P$-induced SC \cite{Hegger,Muramatsu,Knebel,Park}. Previous $^{115}$In-NQR experiments on CeRhIn$_5$ have unraveled a homogeneously coexisting state of AFM and SC in the range $P=1.6-2.1$ GPa. In this novel state of matter, a quantum critical point (QCP) -- the point at which AFM collapses -- and a tetracritical point in the pressure--temperature phase diagram have been found to be around $P_{QCP}=2.1$ GPa and $P_{tetra}$ = 1.98 GPa, respectively \cite{Mito1,SKawasaki,Yashima1}. A neutron diffraction experiment performed at $P$ indicates that the helical structure is maintained at 1.63 GPa \cite{Llobet1}, while other reports state that the $Q_{iz}=0.297$ at $P$ = 0 suddenly changes into  $Q_{iz}\sim 0.4$ near $P=1$ GPa \cite{Majumdar,Raymond}. The CeRh$_{1-x}$Ir$_x$In$_5$ system exhibited a commensurate AFM with $Q_c=(\frac{1}{2},\frac{1}{2},\frac{1}{2})$ in the $P$ range where SC occurs, in addition to exhibiting a helical order similar to that in CeRhIn$_5$. This result suggested an intimate relationship between the onset of SC and the development of a magnetic structure during the coexistence of AFM and SC \cite{Llobet2}. Commensurate AFM was also reported in CeRh$_{1-x}$Co$_{x}$In$_5$ \cite{Yokoyama,Kawamura}. The magnetically ordered Ce moment $M_Q=0.8\mu_B$ at $P=0$ was reported to slightly decrease to $\sim 0.67 \mu_B$ at $P=1.63$ GPa \cite{Llobet1}; a recent experiment \cite{Raymond} reported $M_Q$ = 0.6 and 0.43 $\mu_B$ at $P$= 0 and in the range $1.5-1.7$ GPa, respectively. However, no magnetic structure has so far been observed in the uniformly coexisting phase of AFM and SC in $P$ = $1.6-2.1$ GPa.

In this letter, we report on the $P$-induced evolution of the magnetic structure and its relation to SC through systematic $^{115}$In-NQR studies at the In(1) and In(2) sites in CeRhIn$_5$.  It is revealed that commensurate AFM with $Q_c=(\frac{1}{2},\frac{1}{2},\frac{1}{2})$ is induced above $P\sim$ 1.7 GPa. The present experiments show the strong coupling between AFM and SC OPs.

\section{I\hspace{-.1em}I. Experimental procedure}

The single crystals of CeRhIn$_5$ grown by the self-flux method were moderately crushed into coarse powder in order to allow the RF pulses to easily penetrate the sample for NQR measurements. Hydrostatic pressure was applied by utilizing a NiCrAl-BeCu piston-cylinder cell that was filled with Si-based organic liquid as the pressure-transmitting medium \cite{Kirichenko}. To calibrate the pressure at low temperatures, the shift in $T_c$ of Sn metal at $P$ was monitored by the resistivity measurements. The measurements of the $^{115}$In($I=9/2$)-NQR spectrum  were performed at the transition of 3$\nu_Q$ for the $^{115}$In(2) site in CeRhIn$_5$. CeRhIn$_5$ consists of alternating layers of CeIn and RhIn$_4$, as indicated in Fig. 1a, and there are two sites---In(1) and In(2)---per unit cell. The In(1) and In(2) sites are located in the CeIn layer and RhIn$_4$ layer, respectively. Here, $\nu_Q$ is defined by the NQR Hamiltonian, $\mathcal{H}_Q$ = $(h\nu_Q/6)[3{I_z}^2-I(I+1)+\eta({I_x}^2-{I_y}^2)]$, where $\eta$ is the asymmetry parameter of the electric field gradient ($\nu_Q$ = 16.656 MHz and $\eta$ = 0.445 at the In(2) site, $P=0$ and $T$ = 4.2 K). When an internal magnetic field $H_{int}(\propto M_Q)$ is present at the In sites in association with the onset of AFM, the NQR Hamiltonian is perturbed by the Zeeman interaction, which is given by $\mathcal{H}_{\rm AFM}=- \gamma \hbar\vec{I} \cdot \vec{H}_{int} + \mathcal{H}_Q$. The onset of AFM is signaled by the splitting of the NQR spectrum by $H_{int}$.

%fig1
\begin{figure}[htbp]
\centering
\includegraphics[width=8cm]{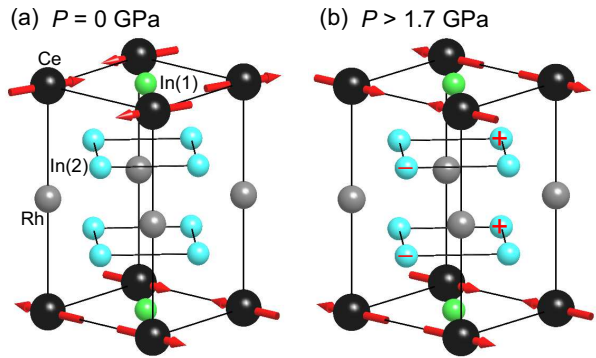}
\caption[]{\footnotesize (color online) Crystal and magnetic structures of CeRhIn$_5$. (a) The incommensurate helical structure along the c-axis with $Q_i=(1/2,1/2,0.297)$ at $P$ = 0. (b) The commensurate antiferromagnetic structure starts to appear under $P$. In this magnetic structure, there are two magnetically different In(2) sites, as shown in this figure. One of these sites considers $H_{int}$ with a ``$+$'' or ``$-$'' sign against the direction of $M_Q$ and the other site considers $H_{int}=0$.  As discussed in reference \cite{Mito2}, the $H_{int}$ at the In(2) site also originates from the direct dipolar field from the Ce magnetic moments and from the indirect ``pseudo-'' dipolar anisotropic field via the supertransferred hyperfine interaction due to the hybridization between the In-5$p$ and Ce-4$f$ orbits.}
\end{figure}

\section{I\hspace{-.1em}I\hspace{-.1em}I. RESULTS AND DISCUSSION}

Figure 2a shows the $P$ dependence of the NQR spectrum (the 3$\nu_Q$ transition) on the In(2) site in the magnetically ordered state of CeRhIn$_5$. Since $M_Q$ rotates about the c-axis with $Q_i=(\frac{1}{2},\frac{1}{2},0.297)$, the amplitude of $H_{int}$ at the In(2) site varies along the axis. As a result, the NQR spectrum at $P=0$ exhibits a very broad pattern (see the spectrum at the bottom of Fig. 2a). This spectrum at $P$ = 0 is well simulated, assuming $H_{int}=H_{max} \cos(2\pi Q_{iz} z)$ as given in reference \cite{Curro}. The solid line in the spectrum for $P=0$ in Fig. 2a represents the distribution $R(\omega)$ of the NQR frequency for $H_{max}=2.5$ kG including an inhomogeneous broadening $\sigma=120$ kHz (Lorentzian function). Here, $R(\omega)=\sum_{\omega_k} f(\omega_k) \frac{\sigma^2}{4(\omega-\omega_k)^2+\sigma^2}$ and $f(\omega)$ is the distribution of the NQR frequency obtained from the distribution of $H_{int}$ through $\mathcal{H}_{\rm AFM}=- \gamma \hbar\vec{I} \cdot \vec{H}_{int} + \mathcal{H}_Q$. However, the NQR spectrum at $P=0$ exhibits a central peak that does not exist in the simulation (solid line), indicating the existence of either a paramagnetic phase or different magnetic phases even at $P=0$. In order to examine the possible contamination of a paramagnetic phase, we present the NQR spectrum (the 1$\nu_Q$ transition) in Fig. 2c at the In(1) site at $T$ = 0.6 K, which is lower than $T_N=3.8$ K. If such a phase exists in the sample, the spectrum shown by the dashed line should be observed. The well-articulated NQR spectrum with two peaks at the In(1) site excludes paramagnetic phases. As demonstrated later, the central peak in the In(2)-NQR spectrum at $P=0$ shown in Fig. 2a arises from commensurate AFM domains shown in Fig. 1b. Incidentally, it is confirmed from the In(2)-NQR spectrum measurements that the volume fraction of the incommensurate AFM changes negligibly below $T_N$ (the complete data are not shown here).

%fig2
\begin{figure}[htbp]
\centering
\includegraphics[width=9cm]{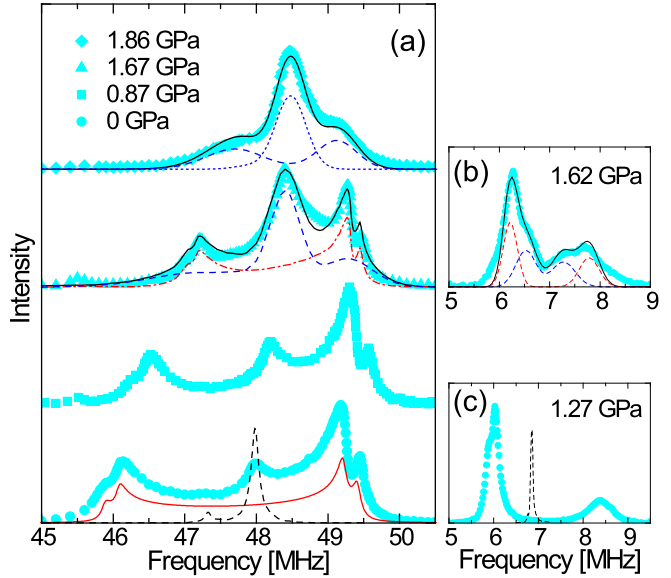}
\caption[]{\footnotesize (color online). (a) The $P$ dependence of the NQR spectrum on the magnetically ordered state in CeRhIn$_5$ at the In(2) site (the 3$\nu_Q$ transition). They were measured at $T$ = 1.5 K, well below $T_N$. The dashed line at $P$ = 0 indicates the NQR spectra of the paramagnetic spectrum measured at $T$ = 4.2 K. At $P$ = 1.67 GPa, the dashed and dotted-dashed lines indicate the simulated spectra for commensurate AFM and the incommensurate helical order, respectively. The solid line represents the convolution of them with fractions of 0.6 and 0.4, respectively. At $P$ = 1.86 GPa, the dashed line represents a simulated spectrum for the commensurate AFM; the two peaks represent the In(2) sites with $H_{int}=\pm 1.1$ kG, which are indicated by ``$+$" and ``$-$" signs in Fig.~1b. The dotted line is the simulated spectrum for the In(2) site with $H_{int}=0$. (b) The spectrum of the 1$\nu_Q$ transition at the In(1) site, $T$ = 0.2 K, and $P$ = 1.62 GPa. The spectrum is well simulated similar to that at In(2) at $P$ = 1.67 GPa in (a). (c) The spectrum of the 1$\nu_Q$ transition at the In(1) site, $T$ = 0.6 K, and $P$ = 1.27 GPa. The spectrum in the paramagnetic state at 4.2 K is indicated by the dashed line.}
\end{figure}

%fig3
\begin{figure}[htbp]
\centering
\includegraphics[width=9cm]{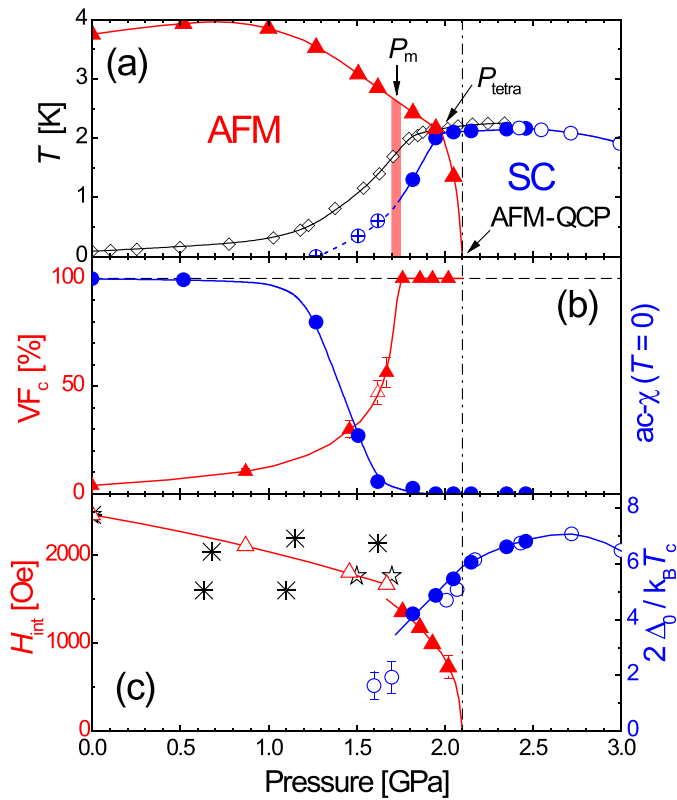}
\caption[]{\footnotesize (color online). (a) The phase diagram of CeRhIn$_5$. $T_N$ (triangles) and $T_c$ (solid circles and encircled crosses) are determined from the In-NQR-$1/T_1$ measurement. The encircled cross indicates $T_c$ for the incommensurate AFM. The $P$ dependence of $T_c$ (diamonds) is determined from the susceptibility measurement performed using a single crystal \cite{Chen}. The open circle indicates $T_c$ obtained from the specific heat measurements \cite{Knebel}. The solid lines are eye guides. $P_{tetra}$ indicates the pressure at the tetracritical point separating AFM, AFM+S, SC, and paramagnetic phases. AFM-QCP points to the quantum critical point at which the AFM collapses. The commensurate AFM is completely realized at $P_m$. (b) The $P$ dependences of VF$_c$ (triangles) and the SC diamagnetism extrapolated to $T$ = 0 (circles). The open triangle at 1.62 GPa is estimated from the analysis of the spectrum at the In(1) site, as shown in Fig. 2b. The solid lines are eye guides. (c) The $P$ dependences of $H_{int}$ on the In(2) sites for the incommensurate AFM ($\vartriangle$) and the commensurate AFM ($\blacktriangle$) along with those of $M_Q$ (\ding{"53} and \ding{"49}) obtained from neutron scattering experiments \cite{Llobet1,Raymond}. $H_{int}$ and $M_Q$ data are normalized at $P$ = 0 ($H_{int}=2.5$ kOe and $M_Q \sim 0.8$(\ding{"53}) or 0.6(\ding{"49}) $\mu_B$). The energy gap $2\Delta_0/k_BT_c$ as a function of $P$ is estimated from a decreasing rate in $1/T_1$ below $T_c$($\bullet$) and the specific-heat jump at $T_c$($\circ$) using the relation $\Delta C = - D(E_F) \; \frac{d \Delta^2 (T)}{dT}|_{T=T_c}$ \cite{Knebel}. Here, the values of $2\Delta_0/k_BT_c$ obtained from the specific heat measurements are normalized as they fit the values obtained from the NQR measurements between 2.1 and 2.5 GPa. The solid lines are eye guides.}
\end{figure}

We focus on a systematic $P$-induced evolution of the NQR spectrum. The NQR spectral shape does not change much at pressures up to 0.87 GPa, except for a slight increase in the intensity of the central peak. As $P$ exceeds 1.6 GPa, however, the NQR intensity at the central peak also increases significantly. Eventually, the spectral shape that is inherent to the helical order disappears at $P=1.86$ GPa (see the spectrum at the top of Fig. 2a). Since paramagnetic phases are excluded from the NQR spectrum at the In(1) site as mentioned above, such a drastic change in the NQR spectrum indicates a $P$-induced change in the AFM structure of CeRhIn$_5$. In fact, the NQR spectral shape at $P=1.86$ GPa is well interpreted by assuming the commensurate AFM with $Q_c=(\frac{1}{2},\frac{1}{2},\frac{1}{2})$ shown in Fig. 1b. In this AFM structure, there are two magnetically different In(2) sites, as shown in Fig. 1b. One of them considers $H_{int}$ with a ``$+$'' or ``$-$'' sign that indicates the up or down direction, respectively, of $\vec{H}_{int}$ along the c-axis at each In(2) site and the other site considers $H_{int}=0$. The characteristic spectrum at $P$ = 1.86 GPa is thus well simulated with $H_{int} = 1.1$ kG, as represented by the solid line in the spectrum at the top of Fig. 2a, thereby demonstrating that the commensurate AFM shown in Fig. 1b occurs near the AFM-QCP in CeRhIn$_5$. It is to be noted that the volume fraction of the commensurate AFM domains (VF$_\mathrm{c}$) increases slightly up to 1 GPa and reaches 100\% at 1.76 GPa, as shown in Fig. 3b. This reveals that the incommensurate AFM completely changes into the commensurate AFM around $P_m \sim 1.7$ GPa. The four peaks in the In(1) spectrum at 1.62 GPa shown in Fig. 2b were reported in a previous paper \cite{Yashima1}. The present NQR measurements suggest that the four peaks are associated with the combination of the commensurate and the incommensurate AFM just below $P_m$. The difference in $H_{int}$ between the commensurate and incommensurate AFM may be due to the difference in the hyperfine coupling. VF$_\mathrm{c}$ is estimated at 1.62 GPa from the analysis of the spectrum at the In(1) site, as shown in Fig. 2b, and it is denoted by an open triangle in Fig. 3b.

We study the $P$-induced evolution of $H_{int}$ at the In(2) site. Fig. 3c indicates that $H_{int}$ progressively decreases from 2.5 kOe at $P$ = 0 to 1.6 kG at $P$ = 1.67 GPa and $M_Q$ decreases from 0.6 $\mu_B$ at $P=0$ to 0.43 $\mu_B$ at $P$ = 1.7 GPa, revealing the relation for $H_{int}^{\rm{In(2)}} \propto M_Q$ in $P$ = $0-1.67$ GPa where the incommensurate domains are dominant. When $P=1.76-2.05$ GPa, where the commensurate AFM is completely established over the entire sample, it is demonstrated that $M_Q$ markedly decreases from 0.36 $\mu_B$ at $P=1.76$ GPa to 0.2 $\mu_B$ at $P$ = 2.02 GPa by using the relation of $H_{int}^{\rm{In(2)}} \propto M_Q$. This result leads us to deduce that at the microscopic level, the SC coexists with the commensurate AFM near the QCP. Fig. 3c shows the $P$ dependence of the SC energy gap $2\Delta_0/k_BT_c$,   which was deduced from the $T$ dependence of $(T_1T)^{-1}$ below $T_c$ using the $d_{x^2-y^2}$-wave model ($\Delta (\theta,\phi)$ = $\Delta_0 \cos 2 \phi$) with parameters $2\Delta_0/k_BT_c$ and the residual density of states at the Fermi level(RDOS). Here, the RDOS is normalized by the density of states at the Fermi level in the PM state.  From the best fitting curves indicated by the solid lines in Fig. 4, we obtained sets of parameters ($2\Delta_0/k_BT_c$, RDOS) = (4.2, 0.42), (5.45, 0.22), and (6.6, 0.14) at 1.82 , 2.05, and 2.35 GPa, respectively. As shown in Fig. 3c, the strong coupling SC is realized above the AFM-QCP in CeRhIn$_5$, which is similar to that in CeCoIn$_5$ \cite{Yashima2}. In CeCoIn$_5$ where the AFM order is absent and the coexistence of AFM and SC is not observed, the RDOS which is mainly derived from the impurity effect is almost constant for $P$ \cite{Yashima2}. However, the RDOS rapidly increases with decreasing $P$ below the AFM-QCP in CeRhIn$_5$, as shown in Fig. 4, suggesting the gapless nature inherent to the coexistence of AFM and SC \cite{SKawasaki,Yashima1}. An important finding presented here is that $2\Delta_0/k_BT_c$ and $T_c$ increase with a significant decrease in $M_Q$ above 1.76 GPa, which provides evidence of strong coupling between the commensurate AFM and SC OPs.

%figt4
\begin{figure}[htbp]
\centering
\includegraphics[width=8cm]{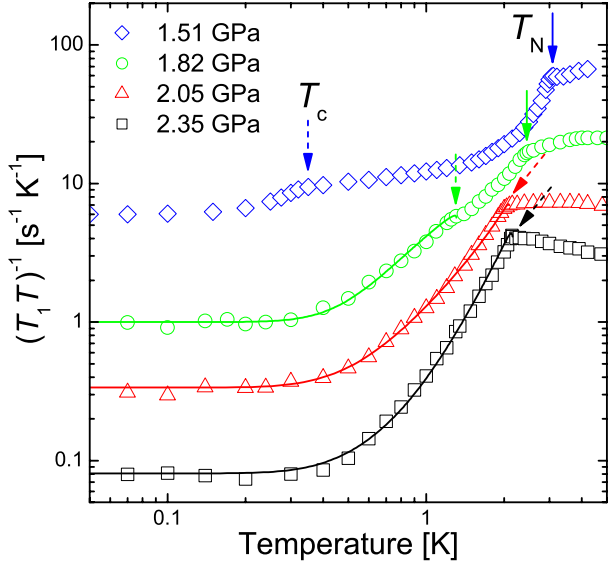}
\caption[]{\footnotesize (color online). The $T$ dependences of $(T_1T)^{-1}$ at pressures above 1.51 GPa. For clarity, note that the $(T_1T)^{-1}$ data at 1.82, 2.05, and 2.35 GPa are divided by 3, 9, and 18, respectively. The solid and dashed arrows indicate $T_N$ and $T_c$, respectively. The solid curves represent fitting curves with parameters $2\Delta_0/k_BT_c$ and RDOS (see text).}
\end{figure}

It is to be noted that the commensurate AFM domains exist even at $P=0$ since the central peak is observed in the NQR spectrum at the bottom of Fig. 2a. This fraction VF$_\mathrm{c}$ is as small as $3-4$\% of the sample used here. However, VF$_\mathrm{c}$ has been confirmed to decrease with an improvement in the quality of the sample. This result suggests that the contamination of the commensurate AFM domains at $P=0$ is possibly associated with the existence of inner stress that is inevitably introduced in the sample, for instance, near the surface region of CeRhIn$_5$. Therefore, it is assumed that the change in the magnetic structure of CeRhIn$_5$ is also sensitive to the existence of strains in the sample introduced by a large pressure gradient in the pressure cell. In one study, Fluorinert-70/77 that freezes at the room temperature near $P$ = 1 GPa was used in neutron experiments performed at $P$ \cite{Sidorov}. The observation of a rise in $Q_z$ from 0.297 to 0.4 at $P$ = 1 GPa may be due to the very large pressure gradient brought about by the application of pressure in the frozen pressure-transmitting medium. It may be difficult to observe commensurate AFM during the neutron scattering measurements since the Ce magnetic moments rapidly decrease above $P_m$.

%fig5
\begin{figure}[htbp]
\centering
\includegraphics[width=8cm]{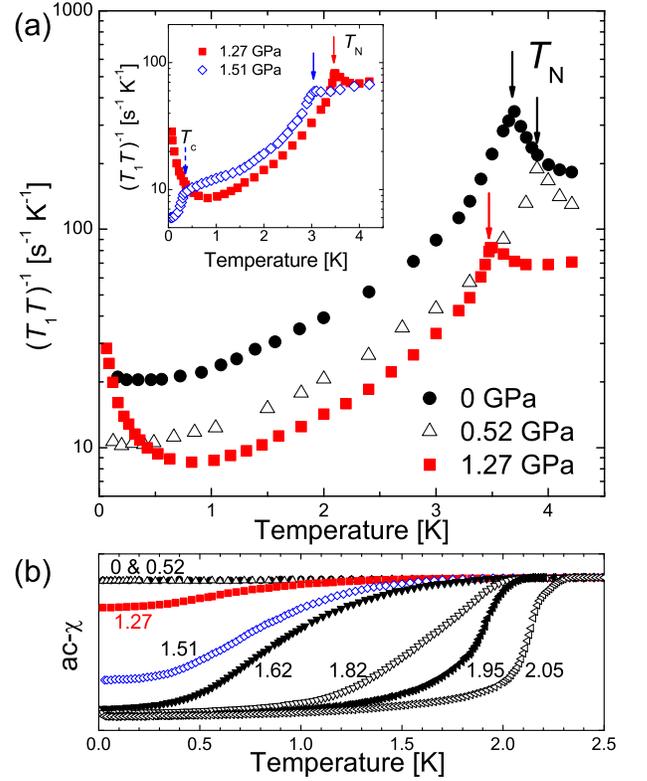}
\caption[]{\footnotesize (color online). (a) The $T$ dependences of $(T_1T)^{-1}$ on the In(1) site for $P$ = 0, 0.52, and 1.27 GPa. The inset shows the $T$ dependences of $(T_1T)^{-1}$ for $P$ = 1.27 and 1.51 GPa. (b) The $T$ dependences of the ac susceptibility between $P$ = 0 GPa and 2.05 GPa obtained by using the in-situ NQR coil. The ac susceptibility is normalized by its value at 2.4 K for each pressure. The fact that the ac-susceptibility at $T=0$ is saturated above $P_m$ suggests that the full SC diamagnetism is realized above $P_m$.}
\end{figure}

It has been demonstrated from the present NQR measurements that SC uniformly coexists with the commensurate AFM and that both OPs are strongly coupled. Another issue is whether the SC can exist in incommensurate AFM. Fig. 5a shows the $T$ dependence of $(T_1T)^{-1}$ below $P_m$ in order to discuss the SC behavior for incommensurate AFM. $(T_1T)^{-1}$ below 0.52 GPa shows a critical slowing down of spin fluctuations at $T_N$ and a $(T_1T)^{-1}$-constant behavior well below $T_N$, representing the characteristics for metallic antiferromagnets. In this $P$ region, the ac-susceptibility does not exhibit SC diamagnetism at all, as shown in Fig. 5b. However, a marked enhancement of $(T_1T)^{-1}$ below 0.6 K and the SC diamagnetism of approximately 16\% were observed at 1.27 GPa, where the incommensurate AFM is still dominant. These results suggest that the superconducting correlation starts to develop around 1.27 GPa, but its effect seems to be confined to a short range since the decrease in $(T_1T)^{-1}$ due to the occurrence of the bulk SC was not found down to the lowest $T$, as shown in Fig. 5a. The enhancement of $(T_1T)^{-1}$ at 1.27 GPa may be induced by SC fluctuations in the incommensurate AFM.

The NQR results mentioned above indicate that there can be two possible scenarios for the SC phenomenon in the incommensurate AFM. One is when the bulk SC appears in the incommensurate AFM. In this case, the SC-QCP exists around 1.27 GPa and $T_c$ continuously starts to increase from 0 K above SC-QCP. However, it is difficult to verify the occurrence of the bulk SC in the incommensurate AFM below $P_m$ since the specific heat jump greatly decreases below 1.7 GPa \cite{Knebel, Knebel2}. If it is true that bulk SC appears even in the incommensurate AFM, theoretical studies are necessary to explain why the specific heat jump at $T_c$ is very small in the coexisting phase of the incommensurate AFM and SC. Another scenario is when the bulk SC does not appear, but instead, the SC correlation is confined to a short range in the incommensurate AFM below $P_m$. It is possible that the incommensurate AFM prevents the development of SC OP. The incommensurate helical magnetic structure, as seen in Fig. 1a, may suppress the growth of the SC correlation along the c-axis in CeRhIn$_5$. As a result, bulk SC does not occur in the incommensurate AFM below $P_m$. Whichever scenario is true, it is remarkable that the SC correlation starts to develop even in the incommensurate AFM regime around 1.27 GPa.

\section{I\hspace{-.1em}V. CONCLUSION}

In conclusion, systematic measurements of the $^{115}$In-NQR spectrum have revealed that commensurate AFM with $Q_c=(\frac{1}{2},\frac{1}{2},\frac{1}{2})$ takes place uniformly over the entire sample above $P_m\sim$ 1.7 GPa near QCP at $P$=2.1 GPa. It is reinforced in the microscopic level that the SC uniformly coexists with the commensurate AFM in such a manner that the magnitude of SC OP and $T_c$ increase as the AFM ordered moment is reduced. This result provides evidence of strong coupling between the commensurate AFM OP and the SC OP. The observation of short-range SC correlation even in the incommensurate AFM regime at 1.27 GPa requires further theoretical studies on the SC characteristics in the incommensurate AFM. The present experiments have shed light on the novel interplay between magnetism and SC in strongly correlated electron systems.

\section{ACKNOWLEDGMENTS}

This work was supported by a Grant-in-Aid for Specially Promoted Research (20001004) and it was also partially supported by the Global COE Program (Core Research and Engineering of Advanced Materials-Interdisciplinary Education Center for Materials Science) from the Ministry of Education, Culture, Sports, Science and Technology (MEXT), Japan. M.Y. was supported by a Grant-in-Aid for Young Scientists (B) of MEXT (20740175).

\end{document}